\documentclass[showpacs,twocolumn,pre]{revtex4}
\usepackage{amssymb,bbm,bm}
\usepackage{graphicx}
\usepackage{bm}
\usepackage{amsmath}
\usepackage[usenames]{color}
\usepackage{epsfig}
\usepackage{hyperref}
\usepackage{subfigure}
\usepackage[sans]{dsfont}
\hypersetup{colorlinks=true, citecolor=blue,
linkcolor=blue,urlcolor=blue }

\begin{document}

\author{Reza Sepehrinia}\email{sepehrinia@ut.ac.ir}
\affiliation{Department of Physics, University of Tehran, Tehran 14395-547,
Iran}

\title{Quantum diffusion of a relativistic particle in a time-dependent random potential}

\begin{abstract}
We present a rigorous study of quantum diffusion of a relativistic particle subjected to a time-dependent random
potential with $\delta$ correlation in time. We find that in the asymptotic time limit the particle wave packet spreads
ballistically in contrast with the nonrelativistic case, which in the same situation exhibits superballistic diffusion.
The relativistic suppression of wave packet diffusion is discussed in connection with statistical conservation laws
that follow from relativistic dynamics.
\end{abstract}


\maketitle

\section{Introduction}

Transport in disordered quantum systems exhibits a variety
of different behaviors ranging from ballistic motion to
diffusion to Anderson localization. In between also lies the
more exotic type known as anomalous diffusion where the
mean squared displacement increases slower than a quadratic
function of time for ballistic motion, but not linearly with
time as in normal diffusion. The other extreme case occurs
when the disorder evolves in time. Temporal fluctuations of
random potential destroy the localization and the transport
rate increases even beyond the ballistic transport. This effect,
which is known as stochastic acceleration in classical physics,
was also predicted in an exactly solvable case of a continuum
Schrodinger equation where it gives rise to cubic growth
of mean squared displacement with time \cite{jayannavar1982nondiffusive}. Recently this
hypertransport has been directly demonstrated in a paraxial
optical experiment \cite{levi2012hyper}. The cubic behavior can also be derived
from classical Langevin dynamics without the dissipative term \cite{jayannavar1982nondiffusive}
which indicates its classical nature. This is expected since
the fluctuations of the potential tend to destroy the quantum
coherence. In fact, the motion of classical particles in a
dynamic random potential exhibits this cubic behavior only
if the potential is $\delta$ correlated in time. In a potential with
finite correlation time the mean squared displacement grows
with different power
$\frac{12}{5}=2.4$ at intermediate time (in one dimension)
\cite{golubovic1991classical,lebedev1995diffusion}, and it was predicted in
Ref. \cite{lebedev1995diffusion} that the true long time asymptotic behavior is
normal diffusion.

Unlike the classical problem spatial correlation of the
potential also has a considerable effect in quantum dynamics.
Even in a rapidly fluctuating potential ($\delta$ correlated in time)
in some cases quantum effects persist for long time \cite{jayannavar1993wave}.
This persistence of quantum effects stems from statistical
conservation laws that govern the quantum dynamics at all
times. These conservation laws are linked to the statistical
properties of the potential. For example, the average kinetic
energy of a quantum particle is conserved for certain types of
spatial correlations of the potential \cite{jayannavar1993wave}, while for a classical
particle it increases linearly with time. As a result of this
conservation, the diffusion of a quantum particle is suppressed
and, instead of the cubic growth, ballistic behavior is observed.

There also exists an earlier analytical study
\cite{madhukar1977exact,madhukar1978exact} of a
discrete model, described with a tight binding Hamiltonian on a lattice,
which leads to normal diffusion. This result is due
to the microscopic length scale of the lattice which induces
a large momentum cutoff. There is no such cutoff in the
continuum model, therefore the particle velocity can increase
indefinitely. On the other hand, indefinite acceleration of the
particle in the continuum model is the result of nonrelativistic
treatment of its motion. So the other feature that might affect
the asymptotic behavior is the relativistic limit on the velocity.
We try to answer the question of how the diffusion of the
wave packet would be influenced by taking the relativistic
effects into account.

Here we adopt the Dirac relativistic wave equation.
Besides the fundamental interest, our motivation has also
been the relevance to various electronic systems which exhibit
properties that can be well described by the Dirac equation.
The main example is graphene, in which electron transport
is essentially governed by the Dirac equation \cite{novoselov2005two}. Many years
after its derivation, the Dirac equation has attracted enormous
attention since the discovery of graphene in the last 10 years.
Two nonintuitive predictions of the Dirac equation, Klein's
paradox and Zitterbewegung, which have been inaccessible
experimentally can now be tested in a condensed matter
experimental setup \cite{katsnelson2006chiral,katsnelson2006zitterbewegung}.
Even the recent progress in
trapped-ion experiments has made it possible to simulate such
relativistic quantum effects \cite{gerritsma2010quantum}.
Narrow-gap semiconductors \cite{zawadzki2005zitterbewegung}, topological
insulators and superconductivity \cite{vafek2014dirac} are other examples in
which the Dirac equation emerges in effective description of quantum dynamics.

\section{Dirac equation and density matrix formalism}

In $1+1$ dimensions which we have $1$ motional degree of freedom the Dirac
equation can be expressed using two Pauli matrices, $\sigma_x$ and $\sigma_z$:
\begin{equation}\label{Dirac equation}
 i\hbar\frac{\partial \psi}{\partial t}=-i\hbar c \sigma_x\frac{\partial \psi}{\partial x}+m c^2 \sigma_z \psi+V(x,t) \psi,
\end{equation}
\begin{widetext}
\noindent where $\psi=\left(
\begin{smallmatrix}
    \psi_1 \\
    \psi_2
\end{smallmatrix}\right)$, $c$ is the speed of light, and $V(x,t)$ is assumed to have a zero average, $\langle V(x,t)\rangle=0$, a translationally invariant correlation
function
\begin{equation}\label{VVcorr}
\langle V(x,t)V(x',t')\rangle=v_0^2\delta(t-t')g(x-x'),
\end{equation}
$\delta$ correlated in time, and an arbitrary correlation in space for
which we will later use specific forms for $g(x)$ for explicit calculations. Here and below,
$\langle \cdots \rangle$ denotes an ensemble average and $\langle \psi|\cdots|
\psi \rangle$ the quantum mechanical expectation value.

We introduce the density matrix
$\bm{\rho}(x',x,t)=\psi^{\dagger}(x',t)\psi(x,t)$, which is itself a $2\times2$
matrix with elements $\rho_{ij}(x',x,t)=\psi_i^{*}(x')\psi_j(x)$, with $i,j=1,2$.
The time evolution $\dot{\bm{\rho}}(x',x,t)=\dot{\psi}^{\dagger}(x',t)\psi(x,t)+\psi^{\dagger}(x',t)\dot{\psi}(x,t)$
is obtained using Eq. (\ref{Dirac equation}) as follows:
\begin{eqnarray}
 &&\dot{\rho}_{11}(x',x,t)=-c\left( \frac{\partial \rho_{21}}{\partial x'}+\frac{\partial \rho_{12}}{\partial x}\right) +\frac{i}{\hbar}[V(x',t)-V(x,t)]\rho_{11}, \\
 &&\dot{\rho}_{22}(x',x,t)=-c\left( \frac{\partial \rho_{12}}{\partial x'}+\frac{\partial \rho_{21}}{\partial x}\right) +\frac{i}{\hbar}[V(x',t)-V(x,t)]\rho_{22}, \\
 &&\dot{\rho}_{12}(x',x,t)=-c\left(\frac{\partial \rho_{11}}{\partial x'}+\frac{\partial \rho_{22}}{\partial x}\right) +\frac{2imc^2}{\hbar}\rho_{12}+\frac{i}{\hbar}[V(x',t)-V(x,t)]\rho_{12}, \\
 &&\dot{\rho}_{21}(x',x,t)=-c\left( \frac{\partial \rho_{11}}{\partial x'}+\frac{\partial \rho_{22}}{\partial x}\right)-\frac{2imc^2}{\hbar}\rho_{21}+\frac{i}{\hbar}[V(x',t)-V(x,t)]\rho_{21}.
\end{eqnarray}
In terms of new variables $\rho=\rho_{11}+\rho_{22}$,
$\sigma=\rho_{11}-\rho_{22}$, $\tau=\rho_{12}+\rho_{21}$,
$\gamma=\rho_{12}-\rho_{21}$, these equations look simpler; also the quantity
$\rho$ is the probability density which is what we need for calculation of
moments of position operator:
\begin{eqnarray}
 &&\dot{\rho}(x',x,t)=-c\left(\frac{\partial}{\partial x}+\frac{\partial }{\partial x'}\right)\tau + \frac{i}{\hbar}[V(x',t)-V(x,t)]\rho, \label{rhodot}\\
 &&\dot{\sigma}(x',x,t)=-c\left(\frac{\partial}{\partial x}-\frac{\partial }{\partial x'}\right)\gamma + \frac{i}{\hbar}[V(x',t)-V(x,t)]\sigma, \\
 &&\dot{\tau}(x',x,t)=-c\left(\frac{\partial}{\partial x}+\frac{\partial }{\partial x'}\right)\rho + \frac{2imc^2}{\hbar}\gamma+\frac{i}{\hbar}[V(x',t)-V(x,t)]\tau, \\
 &&\dot{\gamma}(x',x,t)=-c\left(\frac{\partial}{\partial x}-\frac{\partial }{\partial x'}\right)\sigma + \frac{2imc^2}{\hbar}\tau + \frac{i}{\hbar}[V(x',t)-V(x,t)]\sigma.
\end{eqnarray}
Because the quantities of interest to us are linear in density matrix, we only
need the average of density matrix over random configurations to calculate the
average quantities. To obtain the equations of motion of averaged density
matrix we need to know averages like $\langle V(y,t)\rho(x',x,t)\rangle$.
For a Gaussian random variable $V(x,t)$ we can use the Novikov's identity \cite{novikov1965functionals}:
\begin{equation}\label{meanVrho}
\langle V(y,t)\rho(x',x,t)\rangle=\int dt'' \int dx'' \langle
V(y,t)V(x'',t'')\rangle\left\langle \frac{\delta \rho(x',x,t)}{\delta
V(x'',t'')}\right\rangle.
\end{equation}
The functional derivative in the integrand can be calculated first by integrating Eq. (\ref{rhodot}) with respect to time,
\begin{equation}
\rho(x',x,t)-\rho(x',x,0)=-\int_0^tc\left(\frac{\partial}{\partial
x}+\frac{\partial }{\partial x'}\right)\tau(x',x,t')dt'+
\frac{i}{\hbar}\int_0^t[V(x',t')-V(x,t')]\rho(x',x,t')dt',
\end{equation}
and then
\begin{eqnarray}
\frac{\delta \rho(x',x,t)}{\delta
V(x'',t'')}=&&-\int_{t''}^tc\left(\frac{\partial}{\partial x}+\frac{\partial
}{\partial x'}\right)\frac{\delta \tau(x',x,t')}{\delta V(x'',t'')}dt'+
\frac{i}{\hbar}\int_{t''}^t[V(x',t')-V(x,t')]\frac{\delta \rho(x',x,t')}{\delta
V(x'',t'')}dt'\nonumber \\ &&+
\frac{i}{\hbar}\int_0^t[\delta(x'-x'')-\delta(x-x'')]\delta(t'-t'')\rho(x',x,t')dt'.
\end{eqnarray}
In the first two terms, the lower limit of integrals is changed to $t''$ because the
density matrix depends only on potential at earlier times so the functional
derivatives in the integrands vanish for $t'<t''$. The third term is equal to
$\theta(t-t'')[\delta(x'-x'')-\delta(x-x'')]\rho(x',x,t)$, where $\theta(t)=1$
for $t>0$, $\frac{1}{2}$ for $t=0$, and $0$ for $t<0$. Taking the limit
$t''\rightarrow t$, that we need for Eq. (\ref{meanVrho}), and using Eq. (\ref{VVcorr}),
we have
\begin{eqnarray}
 \langle V(x,t)\rho(x',x,t)\rangle=\frac{i}{2\hbar} v_0^2(g(x-x')-g(0))\langle\rho(x',x,t)\rangle, \\
  \langle V(x',t)\rho(x',x,t)\rangle=\frac{i}{2\hbar} v_0^2(g(0)-g(x-x'))\langle\rho(x',x,t)\rangle.
\end{eqnarray}
By using these quantities we get the following averaged density matrix equations:
\begin{eqnarray}
 &&\frac{\partial}{\partial t}\langle\rho(x',x,t)\rangle=-c\left(\frac{\partial}{\partial x}+\frac{\partial }{\partial x'}\right)\langle\tau\rangle -\frac{v_0^2}{\hbar^2}(g(0)-g(x-x'))\langle\rho(x',x,t)\rangle, \\
 &&\frac{\partial}{\partial t}\langle\sigma(x',x,t)\rangle=-c\left(\frac{\partial}{\partial x}-\frac{\partial }{\partial x'}\right)\langle\gamma\rangle-\frac{v_0^2}{\hbar^2}(g(0)-g(x-x'))\langle\sigma(x',x,t)\rangle, \\
 &&\frac{\partial}{\partial t}\langle\tau(x',x,t)\rangle=-c\left(\frac{\partial}{\partial x}+\frac{\partial }{\partial x'}\right)\langle\rho\rangle + \frac{2imc^2}{\hbar}\langle\gamma\rangle-\frac{v_0^2}{\hbar^2}(g(0)-g(x-x'))\langle\tau(x',x,t)\rangle, \\
 &&\frac{\partial}{\partial t}\langle\gamma(x',x,t)\rangle=-c\left(\frac{\partial}{\partial x}-\frac{\partial }{\partial x'}\right)\langle\sigma\rangle + \frac{2imc^2}{\hbar}\langle\tau\rangle - \frac{v_0^2}{\hbar^2}(g(0)-g(x-x'))\langle\gamma(x',x,t)\rangle.
\end{eqnarray}
With the change of variables $X=\frac12(x+x'), Y=\frac12(x-x'),
\frac{\partial}{\partial X}=\frac{\partial}{\partial
x}+\frac{\partial}{\partial x'},\frac{\partial}{\partial
Y}=\frac{\partial}{\partial x}-\frac{\partial}{\partial x'}$, the
equations are further simplified:
\begin{eqnarray}\label{Rdot}
&& \dot{R}(X,Y,t)=-c\frac{\partial}{\partial X}T(X,Y,t) - h(Y) R(X,Y,t), \\
&& \dot{S}(X,Y,t)=-c\frac{\partial}{\partial Y}G(X,Y,t) - h(Y) S(X,Y,t),\label{Sdot} \\
&& \dot{T}(X,Y,t)=-c\frac{\partial}{\partial X}R(X,Y,t) + \frac{2imc^2}{\hbar} G(X,Y,t) - h(Y) T(X,Y,t), \label{Tdot}\\
&& \dot{G}(X,Y,t)=-c\frac{\partial}{\partial Y}S(X,Y,t) +
\frac{2imc^2}{\hbar} T(X,Y,t) - h(Y) G(X,Y,t), \label{Gdot}
\end{eqnarray}
where $h(Y)=\frac{v_0^2}{\hbar^2} (g(0)-g(2Y))$ and capital Latin letters
represent corresponding Greek letters, e.g.,
$R(X,Y,t)=\langle\rho(x'(X,Y),x(X,Y),t)\rangle$. By applying the Laplace transform
on $t$ and the Fourier transform on $X$ we obtain
\begin{eqnarray}
 && (s+h(Y)) \bar{\tilde{R}}(K,Y,s)=icK \bar{\tilde{T}}(K,Y,s) + \bar{R}_0, \label{Req}\\
 && (s+h(Y)) \bar{\tilde{T}}(K,Y,s)=icK \bar{\tilde{R}}(K,Y,s) + \frac{2imc^2}{\hbar}\bar{\tilde{G}}(K,Y,s) + \bar{T}_0, \label{Teq}\\
 && (s+h(Y)) \bar{\tilde{G}}(K,Y,s)=-c\frac{\partial}{\partial Y} \bar{\tilde{S}}(K,Y,s) +  \frac{2imc^2}{\hbar}\bar{\tilde{T}}(K,Y,s) + \bar{G}_0, \label{Geq}\\
&& (s+h(Y)) \bar{\tilde{S}}(K,Y,s)=-c\frac{\partial}{\partial Y}
\bar{\tilde{G}}(K,Y,s) + \bar{S}_0. \label{Seq}
\end{eqnarray}
Here $R_0$, $T_0$, $G_0$ and $S_0$ are the initial values of the variables (at $t=0$)
which are known if the initial state of particle is determined. For every
variable we use $\tilde{R}(X,Y,s)=\int_0^{\infty}e^{-ts}R(X,Y,t)dt$ and
$\bar{R}(K,Y,t)=\int_{-\infty}^{\infty}e^{iKX}R(X,Y,t)dX$. By eliminating
variables in favor of $\bar{\tilde{G}}$ we get the following second order
differential equation for $\bar{\tilde{G}}$:
\begin{eqnarray}\label{diffeq}
 \frac{\partial^2 \bar{\tilde{G}}}{\partial Y^2}-\frac{h'(Y)}{s+h(Y)}\frac{\partial \bar{\tilde{G}}}{\partial Y}-\frac{1}{c^2}(s+h(Y))^2\left( 1+\frac{4m^2c^4/\hbar^2}{(s+h(Y))^2+c^2K^2}\right) \bar{\tilde{G}}(K,Y,s)=f(K,Y,s),
\end{eqnarray}
where
\begin{eqnarray}
f(K,Y,s)&=&\frac{1}{c}\frac{\partial \bar{S}_0}{\partial Y}-\frac{1}{c}\frac{h'(Y)}{s+h(Y)}\bar{S}_0+\frac{2mc}{\hbar}\frac{ K(s+h(Y))}{(s+h(Y))^2+c^2K^2}\bar{R}_0-\frac{2im}{\hbar}\frac{ (s+h(Y))^2}{(s+h(Y))^2+c^2K^2}\bar{T}_0\nonumber \\
&&-\frac{1}{c^2}(s+h(Y))\bar{G}_0
\end{eqnarray}
and $h'(Y)=dh(Y)/dY$. There is one requirement in order to fix the solution of
this equation and that comes from square integrability of the wave function.
Therefore we seek the solutions that at least vanish at infinity.

\subsection{Kinematics of wave packet}
So far the problem is reduced to solving the differential equation,
Eq. (\ref{diffeq}). The time evolution of the wave packet can be studied through the
moments of probability distribution of the position of the particle. Here we restrict
ourselves to first and second moments (multifractal behavior is also expected
in the higher moments
\cite{bouchaud1992waves,lebedev1995diffusion,jayannavar1993wave}). The moments
of probability distribution $\rho$ can be expressed conveniently as the
derivatives of its Fourier transform in the following way:
\begin{equation}
\mu_n=\langle\langle\psi| \hat{x}^n |\psi\rangle\rangle=\int x^n
\langle\rho(x,x,t)\rangle dx=\left.\frac{1}{i^n}\frac{\partial^n }{\partial
K^n}\bar{R}(K,Y=0,t) \right\vert_{K=0}.
\end{equation}
For $n=0$ we simply get the normalization of the wave function at any time $\langle
\langle \psi|\psi \rangle\rangle = \bar{R}(K=0,Y=0,t)$. Using Eq.
(\ref{Req}) we have $\bar{\tilde{R}}(K=0,Y=0,s)=\frac{\bar{R}_0(K=0,Y=0)}{s}$,
and by the inverse Laplace transform $\langle \langle \psi|\psi
\rangle\rangle=\bar{R}_0(K=0,Y=0)=\langle\psi_0|\psi_0\rangle$, where $\psi_0$
is the initial state. Using the formal expression of $\bar{\tilde{R}}$ in terms
of $\bar{\tilde{G}}$ and initial values, first and second moments are obtained
as follows:
\begin{eqnarray}\label{1stmoment}
 \tilde{\mu}_1 &=& \frac{2imc^3}{\hbar s^2}\bar{\tilde{G}}(K=0,Y=0,s)+\frac{c}{s^2}\bar{T}_0(K=0,Y=0)+\left.\frac{1}{is}\frac{\partial }{\partial K}\bar{R}_0(K,Y=0)\right|_{K=0}, \\
 \tilde{\mu}_2 &=& \left.\frac{4mc^3}{\hbar s^2}\frac{\partial}{\partial K}\bar{\tilde{G}}(K,Y=0,s)\right|_{K=0}+\frac{2c^2}{s^3} \bar{R}_0(K=0,Y=0)-\left.\frac{1}{s}\frac{\partial^2 }{\partial K^2}\bar{R}_0(K,Y=0)\right|_{K=0}. \label{2ndmoment}
\end{eqnarray}

The first term of each of the above equations is unknown and will be determined by solving
Eq. (\ref{diffeq}). We will also be using the properties of the Laplace transform to
determine the asymptotic behavior of the moments, namely, small-$s$ behavior
which corresponds to the long time behavior.
\\
\end{widetext}

\section{Exact solution for free particle with initial Gaussian wave packet}

Evolution of a Gaussian wave packet in the free Schrodinger equation is a
textbook example. An initial Gaussian wave packet, $\left(\frac{2}{\pi
\lambda^2}\right)^{\frac{1}{4}}e^{-\frac{x^2}{\lambda^2}}$ (with zero average
momentum), spreads during the time evolution but always remains Gaussian.
Calculation of the moments of probability density function is therefore simple.
For instance the second moment is given by
$\mu_2=\frac{\lambda^2}{4}+\frac{\hbar^2}{m^2 \lambda^2}t^2$, which exhibits
ballistic spreading of the wave packet. Whereas in general the motion of the
wave packet in the Dirac equation is more complicated. This is due to the existence
of two branches of the energy-momentum relation in the Dirac equation corresponding to
positive- and negative-energy plane wave solutions. Even a initial Gaussian wave
packet, which in general is a superposition of both kinds of plane waves,
becomes non-Gaussian at later times, splits in two parts since positive and
negative components move in opposite direction, and wiggles back and forth
because of the interference of these two parts. These features have already
been discussed using the numerical solution of the Dirac equation
\cite{thaller1992dirac} and time evolution of the position operator in the Heisenberg
picture \cite{barut1981zitterbewegung}. However the time dependence of wave
packets moving according to the Dirac equation usually cannot be determined
explicitly. Here we give the closed expression of the Laplace transform of the density matrix
of a free Dirac particle with an initial Gaussian wave packet.

For a free particle ($v_0\rightarrow 0$) with a normalized initial Gaussian wave
packet,
\begin{equation}\label{}
\psi_0=\psi(x,t=0)=\left(\frac{2}{\pi \lambda^2}\right)^{\frac{1}{4}} \left(
\begin{array}{c}
\alpha \\
\beta
\end{array}\right) e^{-\frac{x^2}{\lambda^2}} , \ \ |\alpha|^2+|\beta|^2=1, \\
\end{equation}
by inserting the corresponding initial density matrix components,
\begin{eqnarray}
&&\bar{R}_0(K,Y)=e^{-\frac{1}{8} \lambda ^2 K^2
}e^{-2\frac{Y^2}{\lambda^2}},  \\
&&\bar{S}_0(K,Y)=(|\alpha|^2-|\beta|^2)e^{-\frac{1}{8} \lambda ^2 K^2
}e^{-2\frac{Y^2}{\lambda^2}}, \\
&&\bar{T}_0(K,Y)=(\alpha^* \beta+\alpha \beta^*)e^{-\frac{1}{8} \lambda ^2 K^2
}e^{-2\frac{Y^2}{\lambda^2}},  \\
&&\bar{G}_0(K,Y)=(\alpha^* \beta-\alpha \beta^*)e^{-\frac{1}{8} \lambda ^2 K^2
}e^{-2\frac{Y^2}{\lambda^2}},
\end{eqnarray}
in Eq. (\ref{diffeq}) and with some algebra, we get
\begin{eqnarray}\label{freediffeq}
&&\frac{\partial^2 \bar{\tilde{G}}}{\partial Y^2}- a^2
\bar{\tilde{G}}=(bY+d)e^{-2\frac{Y^2}{\lambda^2}},
\end{eqnarray}
where $a^2=\frac{1}{c^2}s^2\left( 1+\frac{4m^2c^4/\hbar^2}{s^2+c^2K^2}\right)$
, $b=\frac{-4}{c\lambda^2}(|\alpha|^2-|\beta|^2)e^{-\frac{1}{8} \lambda ^2 K^2
}$, and $d=\frac{2ms/\hbar}{s^2+c^2K^2}(cK+2i\text{Re}(\alpha^*
\beta)s)-2i\text{Im}(\alpha^* \beta)s/c^2$. The general solution for the
corresponding homogeneous equation of Eq. (\ref{freediffeq}) is $C_1e^{a
Y}+C_2e^{-a Y}$, which diverges at both $Y=\infty$ and $Y=-\infty$ ($C_1, C_2$
are constants). Particular solution has also the same asymptotic behavior;
therefore it is possible to determine $C_1, C_2$ by requiring the complete
solution to vanish at infinity. By doing so we obtain the following solution:
\begin{flalign}
\bar{\tilde{G}}= &\frac{\sqrt{\pi}\lambda}{16\sqrt{2} a} e^{\frac{1}{8} a^2
\lambda ^2}\left[\left(a b \lambda^2-4d\right) e^{a Y}
\text{erfc}\left(\frac{\lambda
a}{2\sqrt{2}}+\frac{\sqrt{2}Y}{\lambda}\right)\right. \nonumber\\
& \left.-\left(a b\lambda^2+4d\right)e^{-a Y} \text{erfc}\left(\frac{ \lambda
a}{2\sqrt{2}}-\frac{\sqrt{2}Y}{\lambda}\right)\right].
\end{flalign}
Now we are able to calculate the moments of the probability density of the position.
Using Eqs. (\ref{1stmoment}) and (\ref{2ndmoment}) we have
{
\medmuskip=0mu
\begin{eqnarray}\label{free1stmoment}
\tilde{\mu}_1&=&\frac{2c \text{Re}(\alpha^* \beta) }{ s^2}-\frac{\sqrt{2 \pi }
\lambda  m c^2 e^{\frac{ \lambda ^2 m^2 c^2 }{2 \hbar^2}+\frac{\lambda ^2}{8c^2}s^2}}{s^2 \sqrt{\hbar^2
s^2+4 m^2 c^4}} \nonumber\\
&&\times\left( \frac{2mc^2}{\hbar} \text{Re}(\alpha^*
\beta)+\text{Im}(\alpha^* \beta) s
\right) \nonumber\\
&&\times\text{erfc}\left(\frac{\lambda}{2\sqrt{2}c} \sqrt{s^2+\frac{4 m^2
c^4}{\hbar^2 }}  \right) ,\nonumber\\
\\\label{free2ndmoment}
\tilde{\mu}_2&=&\frac{2 c^2}{ s^3}-\frac{2\sqrt{2 \pi
}\lambda  m^2 c^5e^{\frac{\lambda ^2 m^2 c^2 }{2
\hbar^2}}}{s^3\hbar\sqrt{\hbar^2 s^2+4 m^2 c^4}} \nonumber\\
&&\times\text{erfc}\left(
\frac{\lambda }{2\sqrt{2}c}\sqrt{s^2+\frac{4 m^2 c^4}{\hbar^2 }}
\right)e^{\frac{\lambda ^2}{8c^2}s^2}+\frac{ \lambda ^2}{4 s}.
\end{eqnarray}
}
To see the long time asymptotic behavior of the moments we only need the
behavior of the above expressions at small $s$. Keeping the most divergent terms at
the limit $s\rightarrow 0$ and then by the inverse Laplace transform we obtain
$\mu_1 \approx 2 \text{Re}(\alpha^*\beta)\left(1-\sqrt{\pi}\eta
\text{erfc}(\eta) e^{\eta^2}\right)ct$ and $\mu_2=\left(1-\sqrt{\pi } \eta
\text{erfc} (\eta) e^{\eta^2}\right)c^2 t^2$, where $\eta = \frac{\lambda m
c}{\sqrt{2}\hbar}$.

It is instructive to see how last two results can also be derived from
representation of the position operator in the Heisenberg picture. Using the
Heisenberg equation of motion one finds the velocity operator $\frac{d
\hat{x}}{dt}=\frac{1}{i\hbar}[\hat{x},H_0]=c\sigma_x$, where $H_0=-i\hbar
c\sigma_x\frac{\partial}{\partial x}+mc^2\sigma_z$. For a massive particle the
velocity operator does not commute with the Hamiltonian, so unlike the
nonrelativistic case the velocity is not a constant of motion. The equation of
motion for the velocity can be solved using some operator algebra and then by
integrating the velocity one gets the following result for the position
operator:
{
\medmuskip=0mu
\begin{eqnarray}\label{x(t)}
\hat{x}(t)&=&\hat{x}(0)+\hat{p} c^2 H_0^{-1} t \nonumber \\
&& +  \frac{i}{2}\hbar c
(\sigma_x-\hat{p} c H_0^{-1})H_0^{-1}(e^{-2iH_0t/\hbar}-1),
\end{eqnarray}
}
where $\hat{p}=-i\hbar \partial/\partial x$. The third term is oscillatory in
time and induces the so-called Zitterbewegung. The moments now can be obtained
as the expectation values of different powers of Eq. (\ref{x(t)}) in the initial
wave packet. Especially in the long time limit the first moment is dominated by
the linear term in time and the second moment is dominated by the quadratic term as we
obtained above. The prefactors that we obtained in $\mu_1$ and $\mu_2$ are actually
the expectation values $\langle\psi_0|\hat{p} c^2 H_0^{-1} |\psi_0\rangle$ and
$\langle\psi_0| (\hat{p} c^2 H_0^{-1})^2|\psi_0\rangle$, respectively.

\section{Particle in a time-dependent random potential}

\textit{massless particle}. In the case of a massless particle the density matrix
can be determined exactly too. Because even in the presence of a random potential,
Eqs. (\ref{Req}) and (\ref{Teq}) become decoupled from Eqs.
(\ref{Geq}), (\ref{Seq}) and we have
$\bar{\tilde{R}}(K,Y,s)=\frac{s+h(Y)}{(s+h(Y))^2+c^2K^2}\bar{R}_0+\frac{icK}{(s+h(Y))^2+c^2K^2}\bar{T}_0$.
The function $h(Y)$ vanishes both at $Y=0$ and $v_0=0$; so in the calculation of
moments of position for which we need $\bar{\tilde{R}}$ at $Y=0$, we get the
massless free particle results as discussed earlier. However the effect of
a random potential can be observed in correlations of the wave function at two
different points ($Y\neq 0$). Further with the inverse Laplace and Fourier
transforms we obtain
\begin{eqnarray}\label{massless}
R(X,Y,t)&=&\frac{1}{\sqrt{2\pi}
\lambda}e^{-h(Y)t}([1-2\text{Re}(\alpha^*\beta)]e^{-\frac{2}{\lambda^2}(X+ct)^2}\nonumber\\
&&+[1+2\text{Re}(\alpha^*\beta)]e^{-\frac{2}{\lambda^2}(X-ct)^2}).
\end{eqnarray}
We can see that at later times the wave packet decomposes into two smaller wave
packets moving in opposite directions at the speed of light. This is actually
what happens in the free Dirac equation. Here in addition we have an
exponential factor, due to the random potential, representing the decay of
correlations of the wave function at two points at the distance $Y$ with the time
scale $\tau_0=\frac{1}{h(Y)}$ which depends on the spatial correlation
function. We may also note that this characteristic time diverges both at $Y=0$ and
at the zero disorder limit $v_0=0$.
\begin{figure}[b]
\epsfxsize8truecm \epsffile{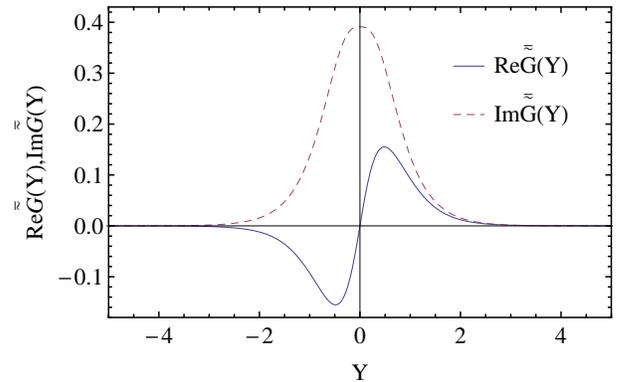} \caption{(Color online) Numerical solution of Eq. (\ref{diffeq})
for $r=1$, $K=0$, $s=0.001$, $\alpha=\sqrt{\frac{4}{5}},\beta=\sqrt{\frac{1}{5}}$ and other parameters equal to $1$.
The real part is odd therefore vanishes at $Y=0$ and the solution is independent of $s$ at small $s$.}\label{G}
\end{figure}

\textit{massive particle}. For this general case the differential equation, Eq.
(\ref{diffeq}), is rather complicated to be solved analytically but the numerical
solution is possible, although imposing the boundary conditions is not straight
forward numerically. As we explained above the solution must vanish at infinity
but numerically we can only determine the solution at a finite interval. At the
boundaries we impose (we use \textsc{mathematica} for numerical solution of differential equation)
a very small value for $\bar{\tilde{G}}$ but we need to make
sure that the interval is large enough so that the result does not depend on
the boundary values. First we note that in order to calculate the moments
(\ref{1stmoment}) and (\ref{2ndmoment}) we only need $\bar{\tilde{G}}$ at $K=0$
and $ \frac{\partial \bar{\tilde{G}} }{\partial K} $ at $K=0$, respectively; so first
we obtain the corresponding equations of these quantities from Eq.
(\ref{diffeq}), which will be relatively simpler. Moreover we are interested in
the long time behavior of moments; therefore we need to know the $s$ dependence of
the Laplace transforms at $s\rightarrow 0$.  We also take the short range spatial
correlation $g(x)=e^{-\left (x/\xi\right )^{2r}}$, where $\xi$ is the
correlation length and $r$ is a positive integer.

Figure \ref{G} shows real and imaginary parts of $\bar{\tilde{G}}(K=0,Y,s)$ for $r=1$, $s=0.01$,
$\alpha=\sqrt{\frac{4}{5}},\beta=\sqrt{\frac{1}{5}}$, and other parameters equal to $1$.
The real part vanishes at $Y=0$, which is necessary because otherwise we obtain
a complex value for $\mu_1$ [see Eq. (\ref{1stmoment})]. Moreover the solution
converges as $s$ goes to $0$; so from Eq. (\ref{1stmoment}) and the inverse Laplace
transform we have $\mu_1 = c\left(1-\frac{2mc^2}{\hbar}
\text{Im}\bar{\tilde{G}}(K=0,Y=0) \right) t$.
\begin{figure}[t]
\epsfxsize8truecm \epsffile{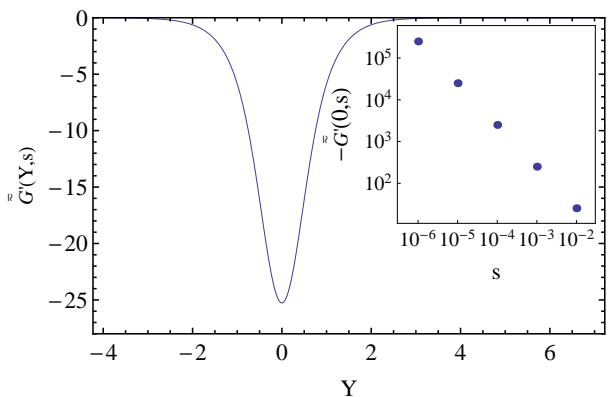} \caption{$\bar{\tilde{G'}}(Y,s)$ for
$r=1$, $s=0.01$ and other parameters equal to $1$. Inset is the plot of
$-\bar{\tilde{G'}}(0,s)$ in log-log scale which shows the $\frac{1}{s}$
divergence in $s\rightarrow 0$.}\label{dG/dK}
\end{figure}
Figure \ref{dG/dK} shows $\bar{\tilde{G'}}(Y,s)=\left.\frac{\partial}{\partial
K}\bar{\tilde{G}}(K,Y,s)\right|_{K=0}$ and the inset is a log-log plot of
$-\bar{\tilde{G'}}(Y=0,s)$ which indicates power law singularity at $s=0$. By fitting the
inset to $s^{-\delta}$ we obtain $\delta=1$, i.e.,
$\bar{\tilde{G'}}(Y=0,s\rightarrow 0)=-\frac{\text{const}}{s}$; therefore the first term in
Eq. (\ref{2ndmoment}) has also $s^{-3}$ singularity at $s=0$ therefore at $t\rightarrow\infty$
\begin{equation}\label{mu2}
\mu_2 \approx
c^2\left(1 + \frac{2mc}{\hbar}\displaystyle \lim_{s\rightarrow 0}
s\bar{\tilde{G'}}(Y=0,s)\right) t^2.
\end{equation}
For comparison the corresponding nonrelativistic result (Eq. (17) of Ref.
\cite{jayannavar1982nondiffusive}), with parameters that used here for $r=1$, is  $\mu_2=\frac{\lambda^2}{4}+
\frac{\hbar^2}{m^2\lambda^2}t^2+\frac{2v_0^2}{3 m^2\xi^2}t^3$.
We repeated the calculations for other values of $r>1$ and obtained the same exponent $\delta=1$
and the same asymptotic behavior of moments.
\section{Summary and Discussion}
To summarize, we have studied the evolution of a wave packet in
the Dirac equation using the density matrix formalism. We derived exactly the
density matrix for a massless particle in a time-dependent random
potential as well as for a free particle with an initial Gaussian wave
packet. For a massive particle in a random time dependent potential we
obtained the asymptotic behavior of average position and mean
squared displacement. In contrast to superballistic diffusion of a
Schrodinger particle for the correlation function $g(x)=e^{-\left (x/\xi\right )^{2r}}$
with $r=1$ [see Eq. (\ref{VVcorr})],
our results exhibit ballistic motion in this case [Eq. (\ref{mu2})]. We find
ballistic behavior in long time limit for $r>1$ as well,
which coincides with nonrelativistic results.

The difference between two cases suggests the existence of another conserved quantity
in the relativistic case. A nonrelativistic quantum particle gains energy from the Gaussian
correlated potential ($r=1$) as its average kinetic energy is not a
constant of motion. For a relativistic particle we have
\begin{widetext}
\begin{eqnarray}
\langle E_K\rangle=\langle \langle \psi |H_0|\psi \rangle \rangle&=&\int dx
\langle \psi^{\dag}(x) H_0 \psi(x) \rangle, \label{KE1}\\&=&\int dx
 \sum_{i,j=1}^{2}\langle\psi_i^{*}(x) (-i\hbar c \sigma_x\partial_x+mc^2 \sigma_z)_{ij} \psi_j(x) \rangle, \label{KE1}\\
&=& \int dx \left[-i\hbar c \frac{\partial}{\partial
x}\langle\tau(x',x,t)\rangle+mc^2\langle\sigma(x',x,t)
\rangle\right]_{x'=x}, \label{KE2}\\
&=& \int dX \left[-i\hbar c \frac{1}{2}
\left(\frac{\partial}{\partial X}+\frac{\partial}{\partial
Y}\right)T(X,Y,t) + mc^2 S(X,Y,t)\right]_{Y=0},\label{KE3} \\
&=& \int dX \left[- \frac{i\hbar c}{2} \frac{\partial}{\partial
Y}T(X,Y,t) + mc^2 S(X,Y,t)\right]_{Y=0}\label{KE4}.
\end{eqnarray}
In going from Eq. (\ref{KE2}) to Eq. (\ref{KE3}) we have performed the change of variables
$X=\frac12(x+x'), Y=\frac12(x-x')$. In the third line, the integral of the term $\frac{\partial}{\partial X} T(X,Y,t)$
vanishes because $T(X=\pm \infty,Y,t)=0$. Also note that
in the integrands the differentiations are being done first and then the values
$x'=x$ ($Y=0$) are set. Now by taking the time derivative and substitution from Eqs. (\ref{Tdot}), (\ref{Gdot})
and using $h(0)=h'(0)=0$, we obtain
\begin{eqnarray}
\frac{d}{dt}\langle E_K\rangle &=& \int dX \left[-\frac{i\hbar c}{2}
\frac{\partial}{\partial Y}\dot{T}(X,Y,t) + mc^2
\dot{S}(X,Y,t)\right]_{Y=0}, \\
&=& \int dX \left[ \frac{i\hbar c^2}{2} \frac{\partial^2R}{\partial X \partial Y} + mc^3\frac{\partial G}{\partial Y}+\frac{i\hbar c}{2}\frac{\partial }{\partial Y}(h(Y)T) \right. \nonumber \\ && \hspace{4.5cm}  \left. - m c^3 \frac{\partial G}{\partial Y}-mc^2 h(Y)S\right]_{Y=0}, \\ &=& \left.\frac{i\hbar c^2}{2} \frac{\partial R}{ \partial Y}\right|_{X=\infty,Y=0}-\left.\frac{i\hbar c^2}{2} \frac{\partial R}{ \partial Y}\right|_{X=-\infty,Y=0}, \\
&=& 0,
\end{eqnarray}
\end{widetext}
which shows that the average kinetic energy is conserved even for the Gaussian correlation
function. The explicit derivation of density matrix that is obtained in
the massless case (Eq. \ref{massless}) reveals an interesting
feature.
The correlation function $\langle\psi^{\dag}(x',t)\psi(x,t)\rangle$, in addition to transient
time dependence due to the motion of the wave front, exhibits an exponential decay
with a time scale $\tau_0=\frac{1}{h\left(\frac{x-x'}{2}\right)}$ induced by
spatial correlations of a time dependent random potential.
\section{aknowledgement}
I would like to acknowledge the hospitality of ICTP where part of
this work was completed.

\bibliography{myref}
\bibliographystyle{apsrev}

\end{document}